\documentclass[12pt]{article}

\usepackage{latexsym,amsmath,amssymb,graphicx}
\usepackage{graphicx}

\evensidemargin \oddsidemargin

\begin{document}


\begin{titlepage}

\renewcommand{\thefootnote}{\fnsymbol{footnote}}



\vspace{15mm}
\baselineskip 9mm
\begin{center}
  {\Large \bf Phase transition of quantum corrected Schwarzschild black hole}
\end{center}

\baselineskip 6mm
\vspace{10mm}
\begin{center}
  Wontae Kim${}$\footnote{Email: wtkim@sogang.ac.kr} and  
  Yongwan Kim${}$\footnote{Email: yongwan89@sogang.ac.kr}
  \\
  \vspace{3mm}
  {\sl ${}$Department of Physics, Sogang University, Seoul, 121-742, Korea\\
  ${}$Center for Quantum Spacetime, Sogang University, Seoul, 121-742, Korea}

\end{center}

\thispagestyle{empty}

\vfill
\begin{center}
{\bf Abstract}
\end{center}
\noindent
We study the thermodynamic phase transition of a quantum-corrected Schwarzschild black hole.  
The modified metric affects the critical temperature which is slightly less than the conventional one.
The space without black holes is not the hot flat space but the hot curved space due to vacuum fluctuations so that
there appears a type of Gross-Perry-Yaffe phase transition even for the very small size of black hole, which
is impossible for the thermodynamics of the conventional Schwarzschild black hole. We discuss 
physical consequences of the new phase transition in this framework.
\\ [5mm]
\vspace*{1cm}
Keywords: Black Hole, Thermodynamics, Phase transition

\vspace{20mm}

\vfill
\end{titlepage}

\baselineskip 6.6mm
\renewcommand{\thefootnote}{\arabic{footnote}}
\setcounter{footnote}{0}

\section{Introduction}
Thanks to Hawking radiation based on a Bekenstein' conjecture \cite{Bekenstein:1973ur, Bekenstein:1974ax},
there has been much attention to the thermodynamics of a black hole system \cite{Hawking:1974sw}.
If the black hole is regarded as a thermal object in equilibrium, then it is natural to apply 
the thermodynamics; however, a crucial difference from the other thermal 
systems is that it is a gravitational object whose entropy is written by the area law \cite{Gibbons:1976ue, Hawking:1982dh} which
provides intriguing thermodynamic issues.

In particular, a hot flat space without black holes can decay into a black hole state because thermal 
particles can be a source of gravitational collapse and then the black hole resides in
thermal equilibrium with the Hawking radiation called Gross-Perry-Yaffe (GPY) phase transition \cite{GPY}. 
From the thermodynamic point of view at the isothermal surface \cite{BCM,SH}, 
one can get a small unstable black hole 
with the mass $M_1$ and a large stable black hole with the mass $M_2$ in the Schwarzschild black hole. 
In connection with the GPY phase transition, the off-shell free energy of the hot flat space without
black holes shows that the GPY phase transition occurs only in the large black hole.  
Actually, the thermodynamic phase transition and behaviors have been well appreciated 
in terms of various ways in the modified Schwarzschild black holes [9-15]. 
To study quantum-mechanical aspects of thermodynamic phase transition,
we have to consider the back reaction of the
spacetime due to quantum fluctuations. In particular, the deformation of the Schwarzschild metric
has been studied in Ref. \cite{KS} for the spherically symmetric
quantum fluctuations of the metric in detail. It may give some improved thermodynamic properties
especially in the UV region although they are expected to be the same with the thermodynamic behaviors at the large black hole.    

In this work, we will study the phase transition of  the quantum-corrected Schwarzschild black hole 
in order to uncover  quantum-mechanical aspects of thermodynamic behaviors. 
On general grounds, the vacuum without black holes at a zero temperature can be defined in terms of the 
Minkowski space. Then, the hot thermal particles in the flat space can decay into black holes.
What it means is that the free energy of the black hole is lower than the free energy  of the hot flat space.
Now, in this quantum-corrected metric, the vacuum without black holes 
is non-trivial since it is not Ricci flat due to quantum fluctuations even in the absence of the black hole. Hence, it is natural to 
regard the hot curved space as an counterpart of the hot flat space for the ordinary Schwarzschild black hole.
As expected, the hot curved space can also decay into the large stable black hole.
For convenience, let us define a tiny black hole whose mass is less than the critical mass, which will be shown in later. Then, even in the UV region, 
we can show that the hot curved space collapses into the tiny black hole.
It can be interpreted as a type of GPY phase transition in the UV region.

In section 2, the quantum-corrected metric given in Ref. \cite{KS} is recapitulated.
The spherically symmetric reduction of the Einstein-Hilbert action can be written in terms of 
a renormalizable two-dimensional dilaton gravity  \cite{RT,ST}, which yields the quantum-corrected metric.
In section 3,   the relevant thermodynamic quantities will be calculated at a finite isothermal surface.
In particular, they vanish at the finite distance before $r = 0$ because of  quantum 
fluctuations. To study the phase transition of the quantum-corrected Schwarzschild black hole,
we construct  the off-shell free energy of the hot curved space and the black hole, and
show that the critical temperature to create the black hole is less than the conventional critical 
temperature in section 4. Moreover, it turns out that 
the free energy of the quantum-corrected black hole is negatively shifted near UV region,
which lies in a lower state than the free energy of the hot curved space. It is the essential ingredient in 
the formation of the tiny black hole.
Finally, the summary and
discussion are given in section 5.

\section{Quantum-corrected Schwarzschild metric}
\label{sec:intro}
In this section, we would like to 
introduce the quantum-corrected metric in a self-contained manner
for our notations  \cite{KS}.
So, we start with the Einstein-Hilbert action with the matter action given by 
\begin{equation}  \label{eq:ac}
I = \int d^4x \sqrt{-g^{(4)}} \left[\frac{R}{16 \pi G_N} +L_{\text{matter}} \right],  
\end{equation}
where $ G_N$ is the Newton constant. From now on, we neglect the classical matter contribution.
Now, the spherically symmetric reduction of the four-dimensional metric can be performed by assuming 
\begin{equation}  \label{eq:ssr}
(ds)_{(4)}^2 =ds_{(2)}^2 + \frac{2G_{N}}{\pi}e^{-2\phi}d\Omega^2 ,
\end{equation}
where we express the radial part in terms of the dilaton field $\phi$ 
maintaining the two-dimensional diffeomorphism. 
Then, we get the two-dimensional dilaton-gravity action \cite{ST}
\begin{equation}  \label{eq:ssrac1}
I = \frac{1}{2\pi} \int d^2x \sqrt{-g^{(2)}} 
\left[e^{-2\phi}R+2 e^{-2\phi}{(\nabla \phi)}^2+\frac{\pi}{G_N}\right].  
\end{equation}
We assume that the generally renormalizable action takes the following form
\begin{equation}  \label{eq:ssrac2}
I = \frac{1}{2\pi} \int d^2x \sqrt{-g} \left[e^{-2\phi}R+2 e^{-2\phi}
{(\nabla \phi)}^2+\frac{\pi}{G_N}U(\phi)\right],   
\end{equation}
with a new general potential $U(\phi)$ for the renormalization.
Next, the divergences can be determined by the two-dimensional nonlinear $\sigma$-model as
\begin{equation}  \label{eq:sigmaac}
I = -\frac{1}{2\pi} \int d^2X \sqrt{-\hat{g}(X)}\left[G_{\alpha \beta}(X)\hat{\nabla} X^{\alpha}\hat{\nabla} X^{\beta}+\frac{1}{2}\Phi(X)\hat{R}+T(X) \right]  
\end{equation}
where $\hat{g}_{\mu \nu}$ is a fiducial metric and $G_{\alpha \beta}(X)$ is a target space metric,
respectively. After the identification of the coordinate $X^\alpha$, the dilaton $\Phi$, the tachyon field T, 
and the target metric $G_{\alpha \beta}(X)$,
one can choose the vanishing $\beta$-function \cite{RT}. Using the renormalization group equation 
for the potential, $ \beta^U=\partial_t U , t=\ln (\mu/\mu_0)$ \cite{KS},
one can get the renormalized potential as
\begin{equation} \label{eq:Uphi}
U(\phi)=\frac{e^{-\phi}}{\sqrt{e^{-2\phi}-\frac{4}{\pi}G_R}} 
\end{equation}
where $G_R=G_N \ln (\mu/\mu_0)$ and $\mu$ is a scale parameter.
Then, solving the equations of motion for the action \eqref{eq:ssrac2},
one can obtain the quantum-corrected Schwarzschild metric,
\begin{align} 
 g(r) &=-\frac{2N}{r}+\frac{1}{r}\int^{r} U(r) dr \nonumber \\
 &=-\frac{2M}{r}+\frac{\sqrt{r^2-a^2}}{r} ,\label{eq:metric}
\end{align}
where $a^2\equiv 4 G_R/\pi$.
The radial coordinate is restricted to $r>a$ and 
the four-dimensional quantum corrected metric is written as
\begin{equation} \label{eq:dsquant}
(ds)^2=-g(r)dt^2+\frac{1}{g(r)}dr^2+r^2d\Omega^2,
\end{equation}
where the event horizon is located at $r_\text{H}=\sqrt{(2M)^2+a^2}$.
\begin{figure}[htp] \label{fig:figrHM}
\center 
\includegraphics[scale=1]{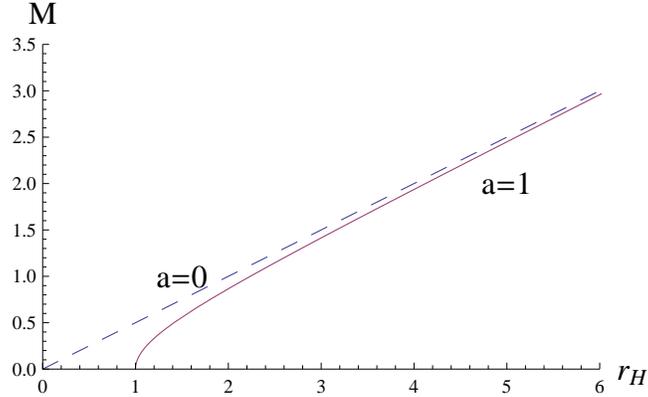} 
\caption{It shows that the modification of relation between the event horizon and the mass.
For a given mass $M$, the size 
of the classical black hole $(a=0)$ is smaller than that of 
the quantum-corrected black hole ($a =1$), which can be seen clearly in the UV region. } 
\end{figure}
Note that the size of the quantum-corrected black hole 
is slightly larger than the classical one as seen from Fig. 1 because of quantum fluctuations.  

The metric \eqref{eq:dsquant} looks asymptotically like a Reisner-Nordstrom metric 
$g(r)\approx 1-2M/r-a^2/2r^2$, however, it gives completely different
behavior because of the negative signature of  the third term in the metric. 
It is interesting to note that the spacetime is not Ricci flat even in spite of the absence of the classical matter contribution,
\begin{align}
R(a) &=\frac{1}{a^2}\left[ 2 \left(\frac{a}{r}\right)^2 \left(1-\frac{1}{\sqrt{1-(\frac{a}{r})^2}}\right)+\left(\frac{a}{r}\right)^4 \left(1-\left(\frac{a}{r}\right)^2 \right)^{-\frac{3}{2}}\right] \label{eq:R}\\
&= \left\{\begin{array}{ll} \infty \qquad & r \to a \\ 0  \qquad & r \to \infty \end{array} \right. \nonumber
\end{align}
where the curvature scalar can be written as asymptotically 
$R\approx 2a^4/r^6 \neq 0$.
The reason why the mass parameter does not appear in the scalar curvature is that 
the original Schwarzschild metric is Ricci flat. The parameter $a$ appears in such a way that
the quantum-mechanical fluctuation breaks the Ricci flatness.
Of course, for the vanishing limit of
$a = 0$, the curvature scalar is zero as expected. 
Essentially, the vacuum fluctuation of the flat spacetime induces the virtual particles, which are the source
of the present curved spacetime.  It means that the vacuum geometry is nontrivial even in spite of the
absence of the black hole $( M = 0)$. 
The classical vacuum corresponding to the flat spacetime was deformed by the spherically
symmetric quantum correction. After all, the ground state is curved. 
From the thermodynamic point of view, if one considers the hot particles in this background, then 
it is natural to consider the instability of the hot curved spacetime, which is an extension of the
Gross-Perry-Yaffe instability of the hot flat spacetime.  

\section{Thermodynamic quantities}
We shall calculate thermodynamic quantities 
in order to study the phase transition from the hot curved space to 
black holes. 
Let us first define the Hawking temperature, 
\begin{equation}
\left. T_H (a)=\frac{1}{4\pi}[\sqrt{-g^{\text{tt}}
g^{\text{rr}}}(-g^{'}_{\text{tt}})]
\right\rvert _{r=r_{\text{H}}}=\frac{1}{4\pi \sqrt{r_{\text{H}}^2-a^2}}.
\end{equation}
It blows up for $r_H = a $.
Next, the observer at the finite isothermal surface sees the Tolman temperature \cite{Tolman, WY} as
\begin{align}
T_{\text{loc}}(a)&=\frac{T_H}{\sqrt{g(r)}} \nonumber \\
		&=\frac{1}{4\pi \sqrt{r_{\text{H}}^2-a^2}} 
\frac{\sqrt{r}}{\sqrt{\sqrt{r^2-a^2}-\sqrt{r_{\text{H}}^2-a^2}}}. \label{eq:TlocrH}
\end{align}
Let us assume that the black hole entropy satisfies the area law, 
\begin{equation} \label{eq:S}
S=\frac{A}{4}=\pi r_{\text{H}}^2,
\end{equation}
which is clear since the present quantum correction just modifies the potential term in the action
so that the area law is consistent with the Wald entropy \cite{Wald}.

The thermodynamic local energy
can be derived from the thermodynamic first law, 
\begin{equation} \label{eq:dE}
dE=TdS,
\end{equation}
which is explicitly calculated as
\begin{align}
E(a)&=E_0+\int_{S_0}^S T_{\text{loc}}(r)dS \nonumber \\
  &=E_0+\sqrt{r}\left[\sqrt{\sqrt{r^2-a^2}}-\sqrt{\sqrt{r^2-a^2}-\sqrt{r_{\text{H}}^2-a^2}}\right], 
\label{eq:E}
\end{align}
using $dS=2\pi r_{\text{H}} dr_{\text{H}}$ in Eq. \eqref{eq:E}.
Note that for $a = 0$, it recovers the well-known local energy of the Schwarzschild black hole.
Specifying the boundary condition of $E_0 =0$, we get $E=M$ for the infinite cavity. In this case, the thermodynamic energy is nothing but the  
ADM mass along with the Hawking temperature so that the thermodynamic
first law $dM=dS/T_{H}$ is trivially satisfied. 
\begin{figure}[htp] \label{fig:figrHC}
\center 
\includegraphics[scale=1]{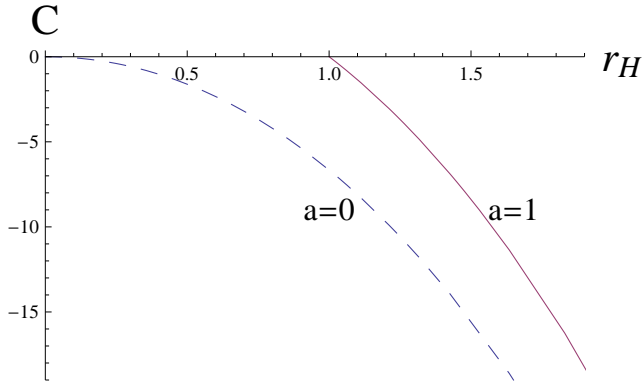} 
\caption{Plot of the heat capacity for $r=10$ at the UV region which is far from the small black hole. 
The solid line for the quantum-mechanical one is slightly shifted and the heat capacity approaches zero 
at the finite size.    } 
\end{figure}

For the thermodynamic stability, 
one can calculate the heat capacity at the finite boundary,
\begin{align}
C(a)&=\left(\frac{dE}{dT_{\text{loc}}}\right)_r \nonumber \\
 &=\frac{4\pi}{3}\frac{(r_{\text{H}}^2-a^2)\left[\sqrt{r^2-a^2}-\sqrt{r_{\text{H}}^2-a^2}\right]}{3\sqrt{r_{\text{H}}^2-a^2}-2\sqrt{r^2-a^2}}. \label{eq:C}
\end{align}
The small black hole is unstable for $r_\text{H}<(2r/3)\sqrt{1+5a^2/4r^2}$ while the
large black hole is stable for $r_\text{H}>(2r/3)\sqrt{1+5a^2/4r^2}$, which is very similar to the conventional Schwarzschild black hole in the box. 
The difference comes from the heat capacity in the UV region so that the vanishing heat capacity for the
quantum-corrected Schwarzschild black hole appears at the finite size as seen from Fig.2. 

As for the Tolman temperature (11) and the heat capacity (15) in connection with the stability of 
the black hole, the Schwarzschild black hole without the box gives rise to thermal instability. 
The essential reason is that Hawking temperature which is measured at the infinity is proportional to the inverse mass, 
so that the Hawking temperature decreases if the black hole absorbs a small amount of radiation.
In other words, it yields the negative heat capacity irrespective of the size of the black hole. 
Moreover, the density of states for the canonical ensemble is pathological because 
it is not well-defined in this black hole system of the negative heat capacity \cite{York:1986it}. 
To overcome these difficulties, one can take the advantage of the Tolman temperature 
by introducing finite thermal bath instead of the infinite thermal bath characterized by the Hawing temperature.
 The Tolman temperature is defined at the surface gravity in terms of the Killing vectors at the finite surface 
so that it contains the red-shift factor of the metric $g(r)$. 
Then, it gives interesting feature that the black hole temperature increases with respect to the mass
in the large black hole for the given size of the cavity, and the heat capacity is eventually positive
 and then the large black hole can be stable. Moreover, the canonical ensemble with the 
Tolman temperautre can be well-defined.

\section{Free energy and phase transition}
We are going to obtain the off-shell free energy to find the critical temperature of the black hole
formation. Then, the phase transition from the hot curved space to the black hole system is studied. 
Now, the off-shell free energy can be defined as
\begin{align}
F_{\text{off}}^{\text{BH}} (a) &=E(a) -TS \nonumber\\     
			 &=\sqrt{r}\left[\sqrt{\sqrt{r^2-a^2}}-\sqrt{\sqrt{r^2-a^2}-2M}\right]-\pi(4M^2+a^2)T. \label{eq:Foff} 
\end{align}
For $a=0$,  it is reduced to the free energy for the Schwarzschild black hole. 
However, the free energy \eqref{eq:Foff}   for $M = 0$ is not zero at any temperatures, 
which is in contrast to the conventional one. It will affect the phase transition from the hot curved space to
the black hole.

\begin{figure}[htp] \label{figMF}
\center
\includegraphics[scale=1]{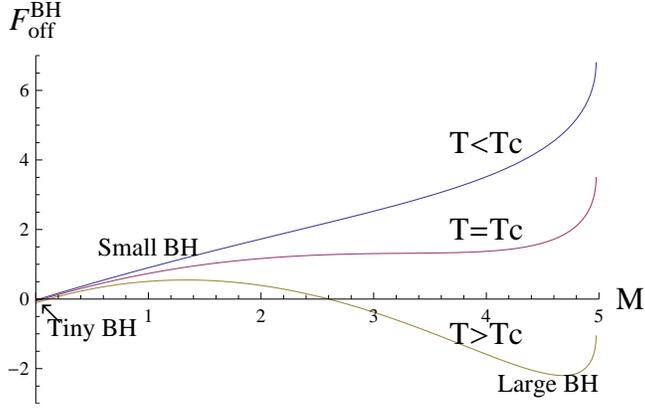}
\caption{For $a=1$, $r=10$,
$T=0.0350 $ and $T_c(1)=0.0205$, the large black hole of mass $M_2 =4.680$ can be nucleated
in stable equilibrium  and the small black hole 
of the mass $M_1=1.330$ can 
 decay into either the large black hole or massless black hole state. 
Overall behaviors are the same with conventional ones except the UV region.  Intriguing thermodynamic properties in the UV region for this tiny black hole
will be given in Fig. 4.}
\end{figure}

By the way, the critical temperature can be calculated as
\begin{equation} \label{eq:Tc}
T_c (a)=\frac{3\sqrt{3}}{8\pi r}\left(1+\left(\frac{a}{r}\right)^2\right)^{-\frac{3}{4}}
\end{equation}
from extrema of the off-shell free energy, $\left.dF_{\text{off}}^{\text{BH}}(a)/dM\right\arrowvert_{T=T_c}=0$.
Among three extrema, the physically meaningful two extrema  in thermal equilibrium appear 
at the positive mass region. 
The small root defined by $M_1(a)$ is for the small unstable black hole and the other one defined
by $M_2(a)$ is for the large stable black hole.  
Note that they are equal root $M_1(a)=M_2(a)$ at the critical temperature. The large black hole can be nucleated above the critical temperature as seen from Fig. 3.
After some calculations, we can find  the small black hole is less than the conventional one while
the large black hole is larger than the conventional one, i.e., 
$M_1(a) < M_1(0)$ and $ M_2(a) > M_2(0) $, where $M_1(0)$ and $M_2(0)$ are just small and large masses for $a =0$. 
In particular,  the quantum-corrected critical temperature is less than the
conventional critical temperature,  which 
means that the large stable black hole can be nucleated in equilibrium at a slightly small temperature compared
to the classically expected temperature.

For the completeness of the phase transition,  we consider the free energy of the hot curved space
at a temperature. For simplicity, the free energy for a single scalar field on the
curved space without black holes is given by  
\begin{align}
F^{\text{HS}}_{\text{off}}(a)&=-\frac{2}{3\pi}\int_a ^r 
dr \frac{r^2}{g(r)} \int_0 ^\infty dE \frac{[E^2-g(r) m^2]^\frac{3}{2}}{(e^{\beta E}-1)}
 \nonumber \\  
&= -\frac{2\pi^3}{135}\sqrt{r^2-a^2}(r^2+2a^2)T^4 +O(m^2).
\end{align}

For a massless limit, it can be regarded as the free energy for gravitons by adding spin degrees of freedom. 
Note that the free energy of the hot flat space is greater than that of the hot curved space.
The reason why we consider the hot curved space rather than the hot flat space is that our spacetime is 
already curved due to the quantum fluctuation, which has something
to do with the non-Ricci flatness of the quantum-corrected Schwarzschild black hole as shown in the previous section.  
In other words, the spacetime without black holes is essentially curved because of
vacuum fluctuations.  The free energy difference between the hot flat space without black hole $F_{\text{off}}^{\text{HS}}(0)$ and 
the hot curved space without black hole
$F_{\text{off}}^{\text{HS}}(a)$  is explicitly given as
\begin{equation}
 F_{\text{off}}^{\text{HS}}(0)-F_{\text{off}}^{\text{HS}}(a)
=-\frac{2\pi^3}{135}r^3 T^4 \left[1-\sqrt{1-\frac{a^2}{r^2}}\left(1+\frac{2a^2}{r^2}\right)\right],\\
\end{equation}
which is positive for $a^2/r^2<\sqrt{3/4}$. If the size of the cavity
is properly large compared to the parameter $a$, the free energy of the hot flat space is greater than the
free energy of 
the  hot curved space. So, one can naturally imagine that the transition from the hot flat space to the 
hot curved space  $F_{\text{off}}^{\text{HS}}(0) \to F_{\text{off}}^{\text{HS}}(a)$ is possible.
\begin{figure}[htp] \label{figMF2}
\center
\includegraphics[scale=1]{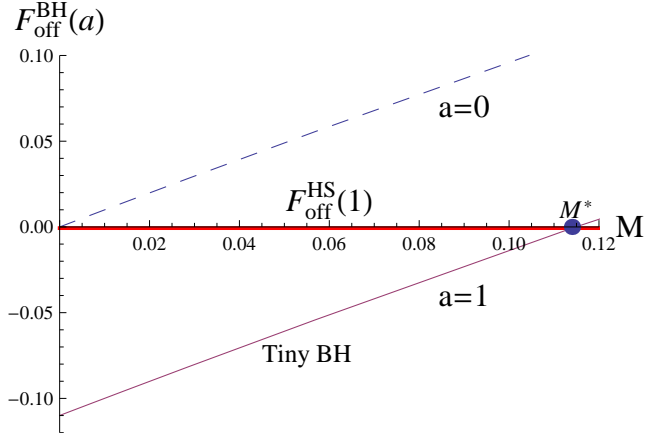}
\caption{Plot of the off-shell free energy at $r=10$,
$T=0.0350$ and $T_c(1)=0.0205$. The horizontal bold line describes the free energy of the hot curved space
which is actually negative. The solid curve
is for the off-shell free energy of the quantum-corrected one $F_{\text{off}}^{\text{BH}}(1)$, which is lower than
the dotted curve of the classical off-shell free energy $F_{\text{off}}^{\text{BH}}(0)$. $M^* =0.114$ is a critical mass to form
a tiny black hole.} 
\end{figure}

We are now in a position to mention the possibility of phase transition using the off-shell free energies
of the hot curved space and the black hole.
In fact, the GPY phase transition for
the Schwarzschild black hole appears only for the 
large black hole, $F_{\text{off}}^{\text{HS}}(0) > F_{\text{off}}^{\text{BH}}(0)$. 
In our case also, the same GPY phase transition
occurs for the large black hole, $F_{\text{off}}^{\text{HS}}(a) > F_{\text{off}}^{\text{BH}}(a)$.   
Moreover, the free energy of the black hole is still lower than the free energy of the hot
curved space even in the UV region,  $F_{\text{off}}^{\text{HS}}(a) > F_{\text{off}}^{\text{BH}}(a)$ 
as long as $T_c (a) < T < [135 a^2/\{2 \pi^2  \sqrt{r^2-a^2}(r^2+2 a^2)\}]^{1/3}$
whereas $F_{\text{off}}^{\text{HS}}(0) < F_{\text{off}}^{\text{BH}}(0)$ for the
conventional case. 
This is plotted in Fig.4 at a temperature greater than the critical temperature.
Note that the mass of the tiny black hole should be   
less than the critical mass $M^* = 0.114$ in Fig.4, which is very small compared to
the mass of the small black hole $M_1=1.330$ in Fig.3.
 Therefore, 
one can see that the hot curved space can be nucleated 
into the tiny black hole; however, it is unstable and
loses its mass eventually.  

\section{Discussions}

We have shown that the phase transition of the quantum-corrected Schwarzschild
black hole is almost the same with the conventional one for the large black hole, which is just the Gross-Perry-Yaffe 
phase transition; however, the critical temperature is less than
that of the Schwarzschild black hole on account of the quantum correction.
In the UV region, the hot curved space without black holes can also decay into the tiny black hole, which means that 
the GPY phase transition occurs
with the help of the quantum correction so that the tiny black hole state 
is more stable than the hot curved space. This tiny black hole is not in thermal equilibrium and subsequently can decay 
into much lower free energy state.

In connection with this state, we would like to mention the end state of the black hole
for $M=0$.
Following the conventional thermodynamic analysis for $T > T_{\text{c}}$,
the energy is zero so that the entropy is naturally zero, which
yields $F_{\text{off}}^{\text{BH}} =0$. However, 
the free energy of the hot flat space is $F_{\text{off}}^{\text{HS}}(0) <0$. 
It means that there does not appear the GPY phase transition.
However, from the beginning, we have considered the quantum-mechanical deformation of the metric 
to explore the UV region because the small size of black holes will receive quantum corrections significantly.
In this case,  the black hole has a minimum size of $r_{\text{H}}=a$ and 
it has a non-vanishing entropy $S=\pi a^2$ with $E=0$. Then, the free energy 
of the black hole becomes negative as $F_{\text{off}}^{\text{BH}}(a)=-\pi a^2 T <0$.
Of course, it is lower than the free energy of the hot curved space.
As a result,  it happens that  $F_{\text{off}}^{\text{HS}}(a) \to F_{\text{off}}^{\text{BH}}(a) \to
\text{ remnant}$ at $M=0$. Although it suggests that there may be some object which has some degrees of
freedom but it is not clear at this
stage in the absence of the full quantized theory.

\section*{Acknowledgments}
We would like to thank E. Choi, M. S. Eune, M. Kim, and Edwin J Son for exciting discussions.
This work was supported by the National Research Foundation of Korea(NRF) 
grant funded by the Korea government (MEST) (2012-0002880) and 
by the National Research Foundation of Korea(NRF) 
grant funded by the Korea government(MEST) through the 
Center for Quantum Spacetime(CQUeST) of Sogang University with grant number 2005-0049409.



\begin{thebibliography}{99}


\bibitem{Bekenstein:1973ur}
  J.~D.~Bekenstein,
  \textit{Black holes and entropy,}
  Phys. Rev. {\bf D7} (1973) 2333.

\bibitem{Bekenstein:1974ax}
  J.~D.~Bekenstein,
  \textit{Generalized second law of thermodynamics in black hole physics,}
  Phys. Rev. {\bf D9} (1974) 3292.

\bibitem{Hawking:1974sw}
  S.~W.~Hawking,
  \textit{Particle Creation by Black Holes,}
  Commun. Math. Phys. {\bf 43} (1975) 199.

\bibitem{Gibbons:1976ue}
  G.~W.~Gibbons, S.~W.~Hawking,
  \textit{Action Integrals and Partition Functions in Quantum Gravity,}
  Phys. Rev. {\bf D15} (1977) 2752.


\bibitem{Hawking:1982dh}
  S.~W.~Hawking, D.~N.~Page,
  \textit{Thermodynamics of Black Holes in anti-De Sitter Space,}
  Commun. Math. Phys. {\bf 87} (1983) 577.


\bibitem{GPY}
D.~J.~Gross, M.~J.~Perry ,and L.~G.~Yaffe,
\textit{Instability of flat space at finite temperature,}
  Phys. Rev. {\bf D25} (1982) 330.

\bibitem{BCM}
J.~D.~Brown, J.~Creighton ,and R.~B.~Mann,
\textit{Temperature, energy, and heat capacity of asymptotically anti-de Sitter black holes,}
  Phys. Rev. {\bf D50} (1994) 6394.


\bibitem{SH}
G.~J.~Stephens and B.~L.~Hu,
\textit{Notes on Black Hole Phase Transitions,}
  Int. J. Theof. Phys. {\bf 40} (2001) 2183.

\bibitem{CFM} 
  G.~Clement, J.~C.~Fabris and G.~T.~Marques,
 \textit{Hawking radiation of linear dilaton black holes,}
  Phys.\ Lett. {\bf B 651} (2007) 54.

\bibitem{BM} 
  R.~Banerjee and B.~R.~Majhi,
 \textit{Quantum Tunneling Beyond Semiclassical Approximation,}
  JHEP {\bf 0806}  (2008) 095.  


\bibitem{KSY-1}
W.~Kim, E.~J.~Son, and M.~Yoon
  \textit{Thermodynamic similarity between the noncummutative Schwarzschild black hole and the Reissner-Nordstr$\ddot{o}$m black hole,}
 JHEP {\bf 04} (2008) 042.


\bibitem{CCO} 
  R.~-G.~Cai, L.~-M.~Cao and N.~Ohta,
  \textit{Black Holes in Gravity with Conformal Anomaly and Logarithmic Term in Black Hole Entropy,}
  JHEP {\bf 1004} (2010) 082.  

\bibitem{RO} 
  M.~E.~Rodrigues and Z.~A.~A.~Oporto,
 \textit{Thermodynamics of phantom black holes in Einstein-Maxwell-Dilaton theory,}
  Phys.\ Rev. {\bf D 85} (2012) 104022.

\bibitem{RJH} 
  M.~E.~Rodrigues, D.~F.~Jardim and S.~J.~M.~Houndjo,
\textit{Thermodynamics of black plane solution,}  arXiv:1205.3481 [gr-qc]. 

\bibitem{JRH} 
  D.~F.~Jardim, M.~E.~Rodrigues and S.~J.~M.~Houndjo,
\textit{ Thermodynamics of phantom Reissner-Nordstrom-AdS black hole,}  arXiv:1202.2830 [gr-qc]. 


\bibitem{KS}
 D.~I.~Kazakov and S.~N.~Solodukhin,
  \textit{On quantum deformation of the Schwarzschild solution,}
  Nucl. Phys. {\bf B429} (1994) 153.


\bibitem{RT} 
  J.~G.~Russo and A.~A.~Tseytlin,
  \textit{Scalar tensor quantum gravity in two-dimensions,} 
 Nucl.\ Phys. {\bf B382} (1992) 259.  

\bibitem{ST} 
  A.~Strominger and S.~P.~Trivedi,
  \textit{Information consumption by Reissner-Nordstrom black holes,}
  Phys.\ Rev. {\bf D48} (1993) 5778.




\bibitem{Tolman}
R. C. Tolman, Phys. Rev. {\bf 35} (1930) 904.

\bibitem{WY} 
  B.~F.~Whiting and J.~W.~York, Jr.,
 \textit{Action Principle and Partition Function for the Gravitational Field in Black Hole Topologies,}
  Phys.\ Rev.\ Lett.\  {\bf 61}  (1988) 1336. 



\bibitem{Wald}
  R.~M.~Wald,
  \textit{Black Hole Entropy is Noether Charge ,}
  Phys.\ Rev. {\bf D48} (1993) 3427.

\bibitem{York:1986it}
  J.~W.~York, Jr.,
  \textit{Black hole thermodynamics and the Euclidean Einstein action ,}
  Phys.\ Rev. {\bf D33} (1986) 2092.





\end{thebibliography}

\end{document}